\documentclass[twocolumn]{article}
\usepackage{epsfig}
\usepackage{graphicx}
\usepackage{color}
\usepackage[acmtitlespace,nocopyright]{acmneww}
\usepackage{times}
\usepackage{amsmath}
\usepackage{amssymb}
\usepackage{xspace}

\usepackage[letterpaper]{geometry}
\usepackage{fullpage}

\newcommand {\mm}[1] {\ifmmode{#1}\else{\mbox{\(#1\)}}\fi}

\newcommand{\denselist}{\itemsep 0pt\parsep=1pt\partopsep 0pt}

\newcommand{\proof}{\noindent{\sc Proof.~}}
\newcommand{\eop}{\hfill\usebox{\smallProofsym}\bigskip}  %
\newsavebox{\smallProofsym}                            
\savebox{\smallProofsym}                               %
{
\begin{picture}(6,6)                                   %
\put(0,0){\framebox(6,6){}}                            %
\put(0,2){\framebox(4,4){}}                            %
\end{picture}                                          %
}                                                      %
\makeatletter
\long\def\@makecaption#1#2{%
  \vskip\abovecaptionskip
  \sbox\@tempboxa{\small #1: #2}%
  \ifdim \wd\@tempboxa >\hsize
    \small #1: #2\par
  \else
    \global \@minipagefalse
    \hb@xt@\hsize{\hfil\box\@tempboxa\hfil}%
  \fi
  \vskip\belowcaptionskip}
\makeatother

\newcommand{\Aspace}        {\mm{{\mathbb A}}}
\newcommand{\Bspace}        {\mm{{\mathbb B}}}
\newcommand{\Mspace}        {\mm{{\mathbb M}}}
\newcommand{\Rspace}        {\mm{{\mathbb R}}}
\newcommand{\Xspace}        {\mm{{\mathbb X}}}
\newcommand{\Yspace}        {\mm{{\mathbb Y}}}
\newcommand{\Zspace}        {\mm{{\mathbb Z}}}
\newcommand{\dime}[1]       {\mm{\rm dim\,}{#1}}

\newcommand{\Pert}          {\mm{\mathcal P}}

\newcommand{\Agroup}        {\mm{\sf A}}
\newcommand{\Bgroup}        {\mm{\sf B}}
\newcommand{\Fgroup}        {\mm{\sf F}}
\newcommand{\Ggroup}        {\mm{\sf G}}
\newcommand{\Hgroup}        {\mm{\sf H}}
\newcommand{\Kgroup}        {\mm{\sf K}}
\newcommand{\Qgroup}        {\mm{\sf Q}}
\newcommand{\Ugroup}        {\mm{\sf U}}
\newcommand{\Vgroup}        {\mm{\sf V}}
\newcommand{\Wgroup}        {\mm{\sf W}}
\newcommand{\Xgroup}        {\mm{\sf X}}
\newcommand{\Betti}         {\mm{\beta}}
\newcommand{\bridge}        {\mm{\cal B}}
\newcommand{\amap}          {\mm{\sf a}}
\newcommand{\bmap}          {\mm{\sf b}}
\newcommand{\fmap}          {\mm{\sf f}}
\newcommand{\jmap}          {\mm{\sf j}}
\newcommand{\umap}          {\mm{\sf u}}
\newcommand{\Ddgm}[1]       {\mm{\rm Dgm}{({#1})}}
\newcommand{\Udgm}[1]       {\mm{\rm Dgm}{({#1})}}
\newcommand{\Rmeasure}[1]   {\mm{\varrho}{({#1})}}

\newcommand{\ti}{\;\;\makebox[0pt]{$\top$}\makebox[0pt]{$\cap$}\;\;}

\newcommand{\diff}[2]       {\mm{\rm D}{#1}{({#2})}}
\newcommand{\rank}[1]       {\mm{\rm rank}{({#1})}}
\newcommand{\kernel}[1]     {\mm{\rm ker\,}{#1}}
\newcommand{\cokernel}[1]   {\mm{\rm cok\,}{#1}}
\newcommand{\image}[1]      {\mm{\rm im\,}{#1}}
\newcommand{\graph}[1]      {\mm{\rm gph\,}{#1}}
\newcommand{\tangent}[2]    {\mm{\rm T}_{#2}{#1}}
\newcommand{\Bdist}[2]      {\mm{W_\infty}{({#1},{#2})}}
\newcommand{\Edist}[2]      {\mm{\|{#1}-{#2}\|}_2}
\newcommand{\Mdist}[2]      {\mm{\|{#1}-{#2}\|}_\Mspace}
\newcommand{\Ydist}[2]      {\mm{\|{#1}-{#2}\|}_\Yspace}
\newcommand{\Maxdist}[2]    {\mm{\|{#1}-{#2}\|}_\infty}
\newcommand{\Pertdist}[2]   {\mm{\|{#1}-{#2}\|}_\Pert}
\newcommand{\seq}[2]        {\mm{F}_{#1}{({#2})}}
\newcommand{\ee}            {\mm{\varepsilon}}
\newcommand{\dd}            {\mm{\delta}}
\newcommand{\capsp}         {{\; \cap \;}}
\newcommand{\cupsp}         {{\; \cup \;}}

\newtheorem{result}{}

\title{Quantifying Transversality ~\\ by Measuring the Robustness of Intersections
       \thanks{This research is partially supported by the Defense
               Advanced Research Projects Agency (DARPA)
               under grants HR0011-05-1-0007 and HR0011-05-1-0057.}
       }

\author{Herbert Edelsbrunner\thanks{Departments of Computer Science
            and of Mathematics, Duke University, Durham, North Carolina,
            Geomagic, Research Triangle Park, North Carolina,
            and IST Austria (Institute of Science and Technology 
Austria).},
        Dmitriy Morozov\thanks{Departments of Computer Science and of
            Mathematics, Stanford University, Stanford, California.}, and
        Amit Patel\thanks{Department of Computer Science,
            Duke University, Durham, North Carolina,
            and IST Austria (Institute of Science and Technology Austria).}
}

\begin{document}
\maketitle

\begin{abstract}
  By definition, transverse intersections are stable under infinitesimal perturbations.
  Using persistent homology, we extend this notion to a measure.
  Given a space of perturbations, we assign to each homology class of the intersection its robustness,
  the magnitude of a perturbation in this space necessary to kill it,
  and prove that robustness is stable.
  Among the applications of this result is a stable notion of robustness
  for fixed points of continuous mappings and a statement of stability for contours
  of smooth mappings.
\end{abstract}

\vspace{0.1in}
{\small
 \noindent{\bf Keywords.}
  Smooth mappings, transversality, fixed points, contours,
  homology, filtrations, zigzag modules, persistence, perturbations, stability.}

\section{Introduction}
\label{sec1}

The main new concept in this paper is a quantification of the classically
differential notion of transversality.
This is achieved by extending persistence from filtrations
of homology groups to zigzag modules of well groups.

\paragraph{Motivation.}
In hind-sight, we place the starting point for the work described in this paper
at the difference between qualitative and quantitative statements and their
relevance in the sciences;
see eg.\ the discussion in Thom's book \cite[Chapters 1.3 and 13.8]{Tho89}.
It appears the conscious mind thinks in qualitative terms, delegating
the quantitative details to the unconscious, if possible.
In the sciences, quantitative statements are a requirement for testing a hypothesis. 
Without such a test, the hypothesis is not falsifiable and,
by popular philosophical interpretation, not scientific \cite{Pop59}.
The particular field discussed in \cite{Tho89} is the mathematical study
of singularities of smooth mappings,
which is dominated by qualitative statements.
We refer to the seminal papers by Whitney \cite{Whi44,Whi55} and the book
by Arnold \cite{Arn92} for introductions.
A unifying concept in this field is the transversality of an intersection
between two spaces.
Its origins go far back in history and appear among others in the work
of Poincar\'{e} about a century ago.
It took a good development toward its present form under Pontryagin, Whitney,
and Thom; see eg.\ \cite{Sma78}.
In his review of Zeeman's book \cite{Zee77},
Smale criticizes the unscientific aspects of the work promoted in
the then popular area of catastrophe theory,
thus significantly contributing to the discussion of qualitative versus
quantitative statements and to the fate of that field.
At the same time, Smale points to positive aspects and stresses
the importance of the concept of transversality in the study of singularities.
In a nutshell, an intersection is transverse if it forms a non-zero angle
and is therefore stable under infinitesimal perturbations;
see Section \ref{sec2} for a formal definition.

\paragraph{Results.}
We view our work as a measure theoretic extension of the
essentially differential concept of transversality.
We extend by relaxing the requirements on the perturbations
from smooth mappings between manifolds
to continuous mappings between topological spaces.
At the same time, we are more tolerant to changes in the intersection.
To rationalize this tolerance, we measure intersections using real numbers
as opposed to $0$ and $1$ to indicate existence.
The measurements are made using the concept of persistent homology;
see \cite{ELZ02} for the original paper.
However, we have need for modifications and use the extension
of persistence from filtrations to zigzag modules as proposed in \cite{CadS08}.
An important property of persistence, as originally defined for filtrations,
is the stability of its diagrams; see \cite{CEH07} for the original proof.
There is no comparably general result known for zigzag modules.
Our main result is a step in this direction.
Specifically, we view the following as the main contributions of this paper:
\begin{enumerate}\denselist
  \item[1.] the introduction of well groups that capture the tolerance
            of intersections to perturbations in a given space of allowable
            perturbations;
  \item[2.] the proof that the diagram defined by the well groups is stable;
  \item[3.] the application of these results to fixed points and periodic orbits of
            continuous mappings.
\end{enumerate}
In addition, our results have ramifications
in the study of the set of critical values,
the apparent contour of a smooth mapping.
Specifically, the stability of the diagrams mentioned above 
results in a stability result for the apparent contour of a smooth mapping from
an orientable $2$-manifold to the plane \cite{EMP09}.
The need for these stable diagrams was indeed what triggered the development
described in this paper.

\paragraph{Outline.}
Section \ref{sec2} provides the relevant background.
Section \ref{sec3} explains how we measure robustness using well groups
and zigzag modules.
Section \ref{sec4} proves our main result, the stability of
the diagrams defined by the modules.
Section \ref{sec5} discusses applications.
Section \ref{sec6} concludes the paper.

\section{Background}
\label{sec2}

We need the algebraic concept of persistent homology to extend the
differential notion of transversality as explained in the introduction.
In this section, we give a formal definition of transversality,
referring to \cite{GuPo74} for general background in differential topology.
We also introduce homology and persistent homology,
referring to \cite{Hat02} for general background in classic algebraic topology
and to \cite{EdHa09} for a text in computational topology.

\paragraph{Transversality.}
Let $\Xspace, \Yspace$ be manifolds,
$f: \Xspace \to \Yspace$ a smooth mapping,
and $\Aspace \subseteq \Yspace$ a smoothly embedded submanifold of the range.
We assume the manifolds have finite dimension and no boundary,
writing $m = \dime{\Xspace}$, $n = \dime{\Yspace}$, and $k = \dime{\Aspace}$.
Given a point $x \in \Xspace$ and a smooth curve $\gamma: \Rspace \to \Xspace$
with $\gamma (0) = x$, we call $\dot{\gamma} (0)$ the \emph{tangent vector}
of $\gamma$ at $x$.
Varying the curve while maintaining that it passes through $x$, 
we get a set of tangent vectors called the
\emph{tangent space} of $\Xspace$ at $x$, denoted as $\tangent{\Xspace}{x}$.
Composing the curves with the mapping,
$f \circ \gamma: \Rspace \to \Yspace$, we get a subset of all smooth curves
passing through $y = f(x) = f \circ \gamma (0)$.
The \emph{derivative} of $f$ at $x$ is
$\diff{f}{x}: \tangent{\Xspace}{x} \to \tangent{\Yspace}{y}$
defined by mapping the tangent vector of $\gamma$ at $x$ to the tangent vector
of $f \circ \gamma$ at $y$.
The derivative is a linear map and its image is a subspace of
$\tangent{\Yspace}{y}$.
The dimensions of the tangent spaces are
$m = \dime{\tangent{\Xspace}{x}}$ and $n = \dime{\tangent{\Yspace}{y}}$,
which implies that the dimension of the image of the derivative
is $\dime{{\diff{f}{x}(\tangent{\Xspace}{x})}} \leq \min \{m,n\}$.

We are interested in properties of $f$ that are stable under perturbations.
We call a property \emph{infinitesimally stable}
if for every smooth homotopy,
$F: \Xspace \times [0,1] \to \Yspace$ with $f_0 = f$,
there is a real number $\dd > 0$ such that $f_t$ possesses the same property
for all $t < \dd$, where $f_t (x) = F(x,t)$ for all $x \in \Xspace$.
An important example of such a property is the following.
The mapping $f$ is \emph{transverse} to $\Aspace$,
denoted as $f \ti \Aspace$,
if for each $x \in \Xspace$ with $f(x) \in \Aspace$,
the image of the derivative of $f$ at $x$ together with the tangent space of
$\Aspace$ at $a = f(x)$ spans the tangent space of $\Yspace$ at $a$.
More formally, $f \ti \Aspace$ if
${\diff{f}{x}(\tangent{\Xspace}{x})} + \tangent{\Aspace}{a} = \tangent{\Yspace}{a}$.
It is plausible but also true that transversality
is an infinitesimally stable property.

\paragraph{Product spaces.}
It is convenient to recast transversality in terms of intersections
of subspaces of $\Xspace \times \Yspace$,
a manifold of dimension $m + n$.
Consider the graphs of $f$ and of its restriction to the preimage of $\Aspace$,
\begin{eqnarray*}
  \graph{f}          &=&  \{ (x,y) \in \Xspace \times \Yspace \mid y = f(x) \} ; \\
  \graph{f|_\Aspace} &=&  \{ (x,a) \in \Xspace \times \Aspace \mid a = f(x) \} .
\end{eqnarray*}
The intersection of interest is between $\graph{f}$ and $\Xspace \times \Aspace$,
two manifolds of dimensions $m$ and $m + k$ embedded in $\Xspace \times \Yspace$.
This intersection is the graph of $f|_\Aspace$, which is homeomorphic to the
preimage of $\Aspace$.
\begin{figure}[hbt]
 \vspace{0.1in}
 \centering
 \resizebox{!}{1.3in}{\input{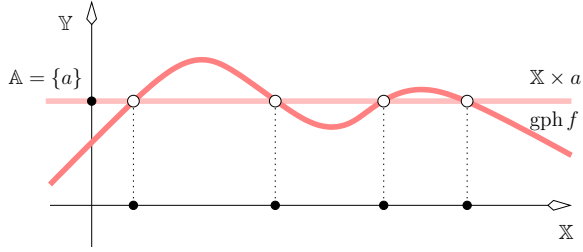}}
 \caption{The preimage of $a$, consisting of four points on the
   horizontal axis representing $\Xspace$, is homeomorphic to the
   intersection of the curve with the horizontal line passing through the
   point $a \in \Yspace$.}
 \label{fig:crossings}
\end{figure}
See Figure \ref{fig:crossings} for an example in which $m = n = 1$ and $k = 0$.
Here, $\tangent{\Aspace}{a} = 0$ and transversality requires that whenever
the curve, $\graph{f}$, intersects the line, $\Xspace \times \Aspace$,
it crosses at a non-zero angle.
This is the case in Figure \ref{fig:crossings} which implies that
having a cardinality four preimage of $a$ is an infinitesimally
stable property of $f$.
Nevertheless, the left two intersection points are clearly more stable than
the right two intersection points,
but we will need some algebra to give precise meaning to
this statement.

\paragraph{Homology.}
The algebraic language of homology is a means to define and count holes
in a topological space.
It is a functor that maps a space to an abelian group
and a continuous map between spaces to a homomorphism between
the corresponding groups.
There is such a functor for each dimension, $p$.
It is convenient to combine the homology groups of all
dimensions into a single algebraic structure.
Writing $\Hgroup_p (\Xspace)$ for the $p$-dimensional homology group of the
topological space $\Xspace$, we form a graded group by taking
direct sums,
\begin{eqnarray*}
  \Hgroup (\Xspace)  &=&  \bigoplus_{p \geq 0} \Hgroup_p (\Xspace) .
\end{eqnarray*}
To simplify language and notation, we will suppress dimensions
and refer to $\Hgroup (\Xspace)$ as the
\emph{homology group} of $\Xspace$.
Its elements are formally written as polynomials,
$\alpha_0 + \alpha_1 t + \alpha_2 t^2 + \ldots$,
where $\alpha_p$ is a $p$-dimensional homology class and only
finitely many of the classes are non-zero.
As usual, adding two polynomials is done componentwise.
The groups $\Hgroup_p (\Xspace)$ depend on a choice of coefficient group.
The theory of persistence introduced below requires we use field coefficients.
An example is modulo two arithmetic in which the field is $\Zspace_2 = \{0, 1\}$.
The $p$-dimensional group is then a vector space,
$\Hgroup_p (\Xspace) \simeq \Zspace_2^{\Betti_p}$,
and its rank, the dimension of the vector space,
is the \emph{$p$-th Betti number}, $\Betti_p = \Betti_p (\Xspace)$.
Similarly, $\Hgroup (\Xspace)$ is a vector space of dimension
$\sum_{p \geq 0} \Betti_p$.
We say $\Xspace$ and $\Yspace$ \emph{have the same homology}
if there is an isomorphism between $\Hgroup (\Xspace)$
and $\Hgroup (\Yspace)$ whose restrictions to the components
are isomorphisms.
Equivalently, $\Betti_p (\Xspace) = \Betti_p (\Yspace)$
for all non-negative integers $p$.

\begin{figure}[hbt]
 \vspace{0.1in}
 \centering
 \resizebox{!}{1.7in}{\input{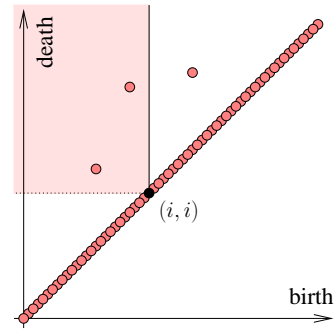}}
 \caption{The three off-diagonal points represent the births and deaths
   of three generators.  The number of points in the shaded upper-left
   quadrant equals the rank of the corresponding homology group.}
 \label{fig:diagram-H}
\end{figure}

\paragraph{Persistent homology.}
Now suppose we have a finite sequence of nested spaces,
$\Xspace_1 \subseteq \Xspace_2 \subseteq \ldots \subseteq \Xspace_\ell$.
Writing $\Phi_i = \Hgroup (\Xspace_i)$ for the homology group
of the $i$-th space, we get a sequence of vector spaces
connected from left to right by homomorphisms induced by inclusion:
$$
  \Phi:  \Phi_1 \to \Phi_2 \to \ldots \to \Phi_\ell .
$$
We call this sequence a \emph{filtration}.
To study the evolution of the homology classes as we progress from left
to right in the filtration,
we let $\varphi_{i,j}$ be the composition of the maps between
$\Phi_i$ and $\Phi_j$, for $i \leq j$.
We say a class $\alpha \in \Phi_i$ is \emph{born} at $\Phi_i$
if it does not belong to the image of $\varphi_{i-1,i}$.
Furthermore, this class $\alpha$ \emph{dies entering} $\Phi_j$ if
$\varphi_{i,j-1} (\alpha)$ does not belong to the image of $\varphi_{i-1,j-1}$
but $\varphi_{i,j} (\alpha)$ does belong to the image of $\varphi_{i-1,j}$.
We call the images of the maps $\varphi_{i,j}$ the
\emph{persistent homology groups} of the filtration
and record the evolution of the homology classes in the
\emph{persistence diagram} of the filtration, denoted as $\Ddgm{\Phi}$.
This is a multiset of points in the extended plane,
$\bar{\Rspace}^2 = (\Rspace \cupsp \{-\infty, \infty\})^2$.
Marking an increase in rank on the horizontal, birth axis
and a drop in rank on the vertical, death axis,
each point represents the birth and the death of a generator
and records where these events happen;
see Figure \ref{fig:diagram-H}.
For technical reasons that will become clear shortly,
we add infinitely many copies of each point on the diagonal to the diagram.
Given an index, $i$, we can read off the rank of
$\Hgroup (\Xspace_i)$ by counting the points in the half-open
upper-left quadrant, $[-\infty, i] \times (i, \infty]$,
anchored at the point $(i, i)$ on the diagonal.
More generally, the rank of the image of $\varphi_{i,j}$ equals
the number of points in the upper-left quadrant
anchored at $(i, j)$.

\paragraph{Stability.}
Consider now the case in which the spaces in the sequence
are sublevel sets of a real valued function $\varphi: \Xspace \to \Rspace$,
that is, there are values $r_i$ such that
$\Xspace_i = \varphi^{-1} (\infty, r_i]$ for each $i$.
A \emph{homological critical value} of $\varphi$ is a value $r$ such that
for every sufficiently small $\dd >0$, the homomorphism
from $\Hgroup (\varphi^{-1} (-\infty, r-\dd])$
to $\Hgroup (\varphi^{-1} (-\infty, r+\dd])$
induced by inclusion is not an isomorphism.
We suppose $\varphi$ is \emph{tame} by which we mean that each sublevel
set has finite rank homology and there are only finitely many homological critical values,
denoted as $r_1 < r_2 < \ldots < r_\ell$.
We can represent the evolution of the homology classes by
the finite filtration consisting of the groups
$\Phi_i = \Hgroup (\varphi^{-1} (-\infty, r_i])$, for $1 \leq i \leq \ell$,
and by the persistence diagram of that filtration, $D = \Ddgm{\Phi}$.
Letting $\psi: \Xspace \to \Rspace$ be another tame function,
we get another filtraton, $\Psi$, and another persistence diagram,
$E = \Ddgm{\Psi}$.
The \emph{bottleneck distance} between the two is the infimum,
over all bijections, $\mu: D \to E$,
of the longest length edge in the matching,
\begin{eqnarray*}
  \Bdist{D}{E}  &=&  \inf_\mu \sup_{a \in D} \Maxdist{a}{\mu(a)} .
\end{eqnarray*}
An important result is the stability of the persistence diagram
under perturbations of the function.
\begin{result}[Stability Theorem for Tame Functions \cite{CEH07}]
  Let $\varphi$ and $\psi$ be tame, real-valued functions on $\Xspace$.
  Then the bottleneck distance between their persistence diagrams is
  bounded from above by $\Maxdist{\varphi}{\psi}$.
\end{result}
Here, $\Maxdist{\varphi}{\psi} = \sup_{x \in \Xspace} |\varphi (x) - \psi (x)|$, as usual.
The original form of this result is slightly stronger as it restricts
itself to dimension preserving bijections.
The theorem implies that the bottleneck distance between the diagrams
defined by $\varphi$ and $\psi$ goes to zero as the difference
between the two functions approaches zero.

\section{Measuring Robustness}
\label{sec3}
The main new concept in this section is the well group defined
by a mapping $f : \Xspace \to \Yspace$, a subspace
$\Aspace \subseteq \Yspace$, and a parameter $r$.
It encodes the persistent homology of the preimage of
the subspace.

\paragraph{Admissible mappings.}
In this paper, we limit the class of mappings to those with manageable properties.
While our goal is a statement of our results in a context
that is sufficiently broad to support interesting applications,
we are aware of the technical burden that comes with generality.
We hope that the following class of mappings gives a happy median between
the conflicting goals of generality and transparency.
\begin{result}[Definition]
  Let $\Xspace$ and $\Yspace$ be topological spaces
  and $\Aspace$ a subspace of $\Yspace$.
  A continuous mapping $f: \Xspace \to \Yspace$ is \emph{admissible}
  if $f^{-1} (\Aspace)$ has a finite rank homology group.
\end{result}
Requiring that the preimage of $\Aspace$ has finite rank homology is
strictly weaker than demanding tameness of the well function defined next.

\paragraph{Perturbations and well groups.}
We are interested in how we explore the neighborhood of
$f^{-1}(\Aspace)$ as we perturb $f$.
For this purpose, we introduce a subspace $\Pert = \Pert(f)$ of 
$C(\Xspace,\Yspace)$, the space of continuous mappings from
$\Xspace$ to $\Yspace$.
For example, $\Pert$ may be the space of continuous mappings $h$
homotopic to $f$;
that is, there exists a continuous mapping $H: \Xspace \times [0,1] \to \Yspace$
with $H(x,0) = f(x)$ and $H(x,1) = h(x)$ for all $x \in \Xspace$.
We assume a metric on $\Pert$, writing $\Pertdist{f}{h}$ for the distance
between two mappings $f, h \in \Pert$.
For example, we could construct one by assuming a metric on $\Yspace$,
lifting it to define the distance between mappings in $C(\Xspace,\Yspace)$,
and taking $\Pertdist{f}{h}$ to be the infimum length of all curves
of mappings that connect $f$ and $h$ within $\Pert$.
%
%
%
We call $h$ an \emph{$r$-perturbation} of $f$ if $\Pertdist{f}{h} \leq r$.

We use these definitions to introduce the \emph{well function}
$f_{\Aspace} \to \Rspace$ of $f$ and $\Aspace$ by setting
$f_{\Aspace}(x)$ to the infimum value of $r$ for which there is an
$r$-perturbation $h \in \Pert$ such that $h(x) \in \Aspace$.
The \emph{level set} of $f_{\Aspace}$ at a value $r$
is the preimage of that value:  $f^{-1}_{\Aspace}(r)$.
The \emph{sublevel set} for the same value $r$ is the union
of the level sets at values at most $r$: $\Xspace_r = f^{-1}_{\Aspace}[0,r]$.
Note that $h_{\Aspace}^{-1}(0) = h^{-1}(\Aspace)$ is contained in $\Xspace_r$
for every $r$-perturbation $h \in \Pert$.
This inclusion induces a homomorphism between the corresponding
homology groups: 
$$
  \jmap_h : \Hgroup (h_\Aspace^{-1} (0)) \to \Fgroup (r),
$$
where we simplify notation by writing $\Fgroup(r)$ for
$\Hgroup (f_{\Aspace}^{-1} [0,r])$.
The image of this map, denoted by $\image{\jmap_h}$,
is a subgroup of $\Fgroup(r)$.
The intersection of subgroups is again a subgroup.
\begin{result}[Definition]
  Given a metric space of perturbations, $\Pert = \Pert(f)$, the
  \emph{well group} of $\Xspace_r$ is the subgroup
  $\Ugroup(r) \subseteq \Fgroup(r)$ obtained by intersecting
  the images, $\image{\jmap_h}$, over all $r$-perturbations
  $h$ of $f$ in $\Pert$. 
\end{result}
A different space of perturbations gives different well groups and therefore
a different interpretation of their meaning.

\paragraph{Example.}
To illustrate the definitions, let us consider again the example in
Figure \ref{fig:crossings} of the mapping from the real line to itself.
The preimage of $\Aspace = \{a\}$ is a set of four points
separated by three critical points of $f$.
From left to right, the values of $f$ at these critical points are
$a + r_1$, $a - r_2$, $a + r_3$.
Correspondingly, the well function, $f_a : \Rspace \to \Rspace$,
has three homological critical values,
namely $r_1 > r_2 > r_3$.
Setting $\Pert = \Pert(f)$ to the set of all continuous mappings 
$h : \Rspace \to \Rspace$ and measuring distance by
$\Pertdist{f}{h} = \sup_x |f(x)-h(x)|$,
we have a well group $\Ugroup(r)$ for each radius $r \geq 0$.
Table \ref{tbl:ranks} shows the ranks of $\Fgroup (r)$ and $\Ugroup (r)$
for values of $r$ in the interior of the four intervals delimited by the homological critical values.
\begin{table}[hbt]
 \vspace*{0.1in}
 \centering
 \small
 \begin{tabular}{c|cccc}
                  &  $(0,r_3)$  &  $(r_3,r_2)$  &  $(r_2,r_1)$  &  $(r_1,\infty)$  \\ \hline
   $\Fgroup (r)$  &      4      &       3       &       2       &        1         \\
   $\Ugroup (r)$  &      4      &       2       &       2       &        0
 \end{tabular}
 \caption{The ranks of the homology and well groups defined for the mapping $f$
   and the submanifold $\Aspace = \{a\}$ in Figure \protect{\ref{fig:crossings}}.}
 \label{tbl:ranks}
\end{table}
Starting with $r = 0$, we have four points, each forming a component
represented by a class in the homology group and in the well group
of the sublevel set of $f_a$.
Therefore, both groups are the same and have rank four,
see the first column in Table \ref{tbl:ranks}.
Growing $r$ turns the points into intervals but leaves the groups
the same until $r$ passes $r_3$,
the smallest of the three homological critical values.
The two right intervals merge into one,
so the rank of the homology group drops to three.
We can find an $r$-perturbation, $r > r_3$, whose level set at $a$ consists
of the left two points of $f^{-1} (a)$ but the right two points
have disappeared.
Indeed, the level set of every $r$-perturbation, for $r  > r_3$, has a
non-empty intersection with the first two but can have empty intersection
with the merged interval on the right.
Hence, the left two intervals have a representation in the well group,
the merged interval does not, and the rank of the well group is two;
see the second column in Table \ref{tbl:ranks}.
The next change happens when $r$ passes $r_2$.
The middle interval merges with the merged interval
on the right.
The rank of the homology group drops to two, while the rank of
the well group remains unchanged at two;
see the third column in Table \ref{tbl:ranks}.
Finally, when $r$ passes $r_1$, the remaining two intervals merge
into one, so the rank of the homology group drops to one.
We can find an $r$-perturbation, $r > r_1$, whose level set at $a$
is empty, so the rank of the well group drops to zero;
see the last column in Table \ref{tbl:ranks}.

\paragraph{Terminal critical values.}
Recall that we assume the mapping $f: \Xspace \to \Yspace$ is admissible.
The initial homology group, $\Fgroup (0) = \Hgroup (f_\Aspace^{-1} (0))$,
has therefore finite rank,
and because $\Ugroup (0) \subseteq \Fgroup (0)$,
the initial well group has finite rank.
For convenience, we permit negative parameter values by stipulating
$\Fgroup(r) = \Fgroup(0)$ and $\Ugroup(r) = \Ugroup(0)$
whenever $r \leq 0$.
Imagine we grow the sublevel set by gradually increasing $r$ to infinity.
Since the admissibility of $f$ does not imply the tameness of the well function,
this leaves open the possibility that $f_\Aspace$ has infinitely many homological
critical values.
We call a radius, $r$, a \emph{terminal critical value} of $f_\Aspace$
if for every sufficiently small $\dd > 0$ the homomorphism
from $\Fgroup (r-\dd)$ to $\Fgroup (r+\dd)$ applied to $\Ugroup (r-\dd)$ does not
give $\Ugroup (r+\dd)$.
In contrast to the homological critical values, there can only be a finite number
of terminal critical values.
To see this, we note that the set of perturbations that define the well groups
grows with increasing $r$.
It follows that the well groups can not increase in rank.
To state this relationship between well groups more formally,
we write $\fmap (r,s) : \Fgroup (r) \to \Fgroup (s)$
for the homomorphism induced by inclusion.
\begin{result}[Shrinking Wellness Lemma]
  For each choice of radii $r \leq s$, the image of the well group at $r$
  contains the well group at $s$, that is,
  $\Ugroup (s) \subseteq \fmap (r,s) (\Ugroup (r))$.
\end{result}
The only way the well group can change is by lowering its rank.
Since we start with a finite rank,
there can only be finitely many terminal critical values,
which we denote as $u_1 < u_2 < \ldots < u_l$.
To this sequence, we add $u_0 = -\infty$ on the left and $u_{l+1} = \infty$ on the right.
We choose an interleaved sequence 
$$
  u_0 < r_0 < u_1 < \ldots < u_{l} < r_{l} < u_{l+1}
$$
and index the homology and the well groups accordingly,
writing $\Fgroup_i = \Fgroup (r_i)$ and $\Ugroup_i = \Ugroup (r_i)$, for all $i$.
To these sequences, we add $\Fgroup_{-1} = \Ugroup_{-1} = 0$ on the left
and $\Fgroup_{l+1} = \Ugroup_{l+1} = 0$ on the right.
Furthermore, we write $\fmap_{i,j} : \Fgroup_i \to \Fgroup_j$
instead of $\fmap(r_i,r_j)$ for all feasible choices of $i \leq j$.

\begin{figure}[hbt]
 \vspace{0.1in}
 \centering
 \resizebox{!}{1.3in}{\input{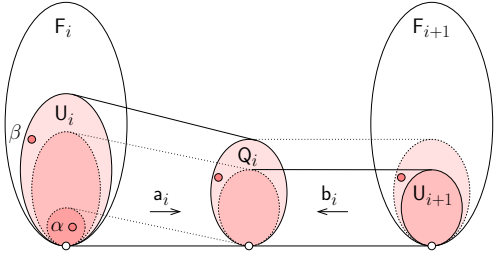}}
 \caption{Connecting two consecutive well groups to the
   quotient group introduced between them.
   The class $\alpha$ dies a conventional death and the class $\beta$
   dies an unconventional death.}
 \label{fig:well}
\end{figure}

\paragraph{Well module.}
In contrast to the homology groups, the well groups of the sublevel sets
do not form a filtration.
Instead, they form a special kind of zigzag module.
By definition of terminal critical values,
the rank of $\Ugroup_i$ exceeds the rank of $\Ugroup_{i+1}$.
The rank of the image, $\fmap_{i,i+1} (\Ugroup_i)$,
is somewhere between these two ranks.
We call a difference between $\Ugroup_i$ and its image a
\emph{conventional death}, in which a class maps to zero,
and a difference between the image and $\Ugroup_{i+1}$
an \emph{unconventional death}, in which the image of a class
lies outside the next well group.
We capture both cases by inserting a new group between the
contiguous well groups; see Figure \ref{fig:well}.
To this end, we consider the restriction of $f_{i,i+1}$ to $\Ugroup_i$
and in particular its kernel, $\Kgroup_i = \Ugroup_i \capsp \kernel{f_{i,i+1}}$,
which we refer to as the \emph{vanishing subgroup} of $\Ugroup_i$.
Using this subgroup, we construct $\Qgroup_i = \Ugroup_i / \Kgroup_i$.
The forward map, $\amap_i: \Ugroup_i \to \Qgroup_i$,
is defined by mapping a class $\xi$ to $\xi + \Kgroup_i$.
It is clearly surjective.
The backward map, $\bmap_i: \Ugroup_{i+1} \to \Qgroup_i$,
is defined by mapping a class $\eta$ to $\xi + \Kgroup_i$,
where $\xi$ belongs to $\fmap_{i,i+1}^{-1} (\eta)$.
This map is clearly injective.
Instead of a filtration in which all maps go from left to right,
we get a sequence in which the maps alternate between going forward and backward.
As indicated below, every other group in the sequence
is a subgroup of the corresponding homology group:
$$
  \begin{array}{ccccccccc}
    \Qgroup_{i-1}  &  \stackrel{\bmap_{i-1}}{\leftarrow}   &  \Ugroup_i
                   &  \stackrel{\amap_{i}}{\rightarrow}    &  \Qgroup_i
                   &  \stackrel{\bmap_{i}}{\leftarrow}     &  \Ugroup_{i+1}
                   &  \stackrel{\amap_{i+1}}{\rightarrow}  &  \Qgroup_{i+1}     \\
                 & & \downarrow & &             & & \downarrow     & &         \\
    \rightarrow  & & \Fgroup_i  & & \rightarrow & & \Fgroup_{i+1}  & & \to .
  \end{array}
$$
We call this sequence the \emph{well module} of $f_\Aspace$,
denoted as $\Ugroup$.
We remark that $\Ugroup$ is a special case of a zigzag module as introduced in \cite{CadS08}.
It is special because all forward maps are surjective and all backward maps are injective.
Equivalently, there are no births other than at the start,
when we go from $\Ugroup_{-1}$ to $\Ugroup_0$.

\begin{figure}[hbt]
 \vspace{0.1in}
 \centering
 \resizebox{!}{1.4in}{\input{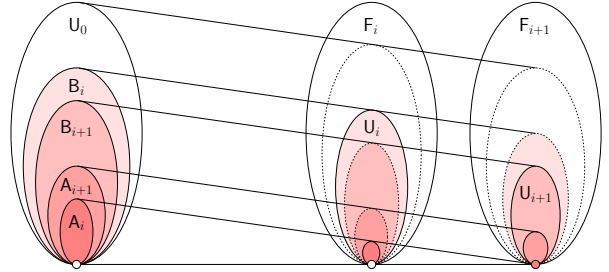}}
 \caption{The left filtration decomposes $\Ugroup_0$ into the preimages of the
   well groups and the preimages of their vanishing subgroups.}
 \label{fig:left-filtration}
\end{figure}

\paragraph{Left filtration.}
Perhaps surprisingly, the evolution of the homology classes can still be
fully described by pairing births with deaths, just like for a filtration.
To shed light on this construction, we follow \cite{CadS08} and turn a
zigzag module into a filtration.
In our case, all births happen at $\Ugroup_0$,
so this transformation is easier than for general zigzag modules.
Write $\umap_{0,i}: \Ugroup_0 \to \Fgroup_i$ for the restriction of
$\fmap_{0,i}$ to the initial well group.
By the Shrinking Wellness Lemma, the image of this map contains
the $i$-th well group, that is,
$\Ugroup_i \subseteq \umap_{0,i} (\Ugroup_0)$.
We consider the preimages of the well groups in $\Ugroup_0$ together
with the preimages of their vanishing subgroups,
$\Agroup_i = \umap_{0,i}^{-1} (\Kgroup_i)$ and $\Bgroup_i = \umap_{0,i}^{-1} (\Ugroup_i)$;
see Figure \ref{fig:left-filtration}.
We note that $\Agroup_i / \Agroup_{i-1} \simeq \kernel{\amap_i}$ and
$\Bgroup_i / \Bgroup_{i+1} \simeq \cokernel{\bmap_i}$.
In words, the first quotient represents the homology classes that die
a conventional death, going from $\Ugroup_i$ to $\Ugroup_{i+1}$,
and the second quotient represents the homology classes
that die an unconventional death in the same transition.
As illustrated in Figure \ref{fig:left-filtration},
the preimages form a nested sequence of subgroups of $\Ugroup_0$.
Together with the inclusion maps, this gives the
\emph{left filtration} of the zigzag module,
$$
  0 \to \Agroup_0 \to \ldots \to \Agroup_{l} = \Bgroup_{l} \to \ldots \to \Bgroup_0 = \Ugroup_0 .
$$
We can recover the well groups with $\Ugroup_i \simeq \Bgroup_i / \Agroup_{i-1}$.
Recall that $\Ugroup_{l+1} = 0$, which implies $\Kgroup_{l} = \Ugroup_{l}$.
It follows that the middle two groups in the left filtration,
$\Agroup_{l}$ and $\Bgroup_{l}$, are indeed equal.

\paragraph{Compatible bases.}
A useful property of the left filtration is the existence of compatible
bases of all its groups.
By this we mean a basis of $\Ugroup_0$ that contains a basis for
each $\Agroup_i$ and each $\Bgroup_i$.
Specifically, we rewrite $\Ugroup_0$ as a direct sum of kernels
of forward maps and cokernels of backward maps:
\begin{eqnarray*}
  \Ugroup_0  &\simeq&  \kernel{\amap_0} \oplus \cdots \oplus \kernel{\amap_{l}} \oplus \\
             &      &  \cokernel{\bmap_{l-1}} \oplus \cdots \oplus \cokernel{\bmap_0}.
\end{eqnarray*}
Reading this decomposition from left to right, we encounter
the $\Agroup_i$ and the $\Bgroup_i$ in the sequence they occur
in the left filtration.
Choosing a basis for each kernel and each cokernel,
we thus get compatible bases for all groups in the left filtration.
We call this the \emph{left filtration basis} of $\Ugroup_0$.
It is unique up to choosing bases for the kernels and cokernels.

Consider now a homology class $\alpha$ in $\Ugroup_0$
and its representation as a sum of basis vectors.
We write $\alpha (\amap_i)$ for the projection of $\alpha$ to
the preimage of the kernel
of the $i$-th forward map, which is obtained by removing all vectors
that do not belong to the basis of $\kernel{\amap_i}$.
Similarly, we write $\alpha (\bmap_i)$ for the projection of $\alpha$
to the preimage of $\cokernel{\bmap_i}$.
Letting $j$ be the minimum index such that $\alpha(\amap_i) = \alpha(\bmap_i) = 0$
for all $i \geq j$,
we say that $\alpha$ \emph{falls ill} at $\Ugroup_j$.

\paragraph{Well diagrams.}
Constructing the birth-death pairs that describe the well module is now easy.
All classes are born at $\Ugroup_0$, however,
to distinguish the changes in the well group from those in the homology group,
we say all the classes \emph{get well} at $\Ugroup_0$.
They fall ill later, and once they fall ill, they do not get well any more.
The drop in rank from $\Ugroup_{i-1}$ to $\Ugroup_i$
is $\mu_i = \rank{\kernel{\amap_{i-1}}} + \rank{\cokernel{\bmap_{i-1}}}$.
We thus have $\mu_i$ copies of the point $(0, u_i)$ is the diagram.
There is no information in the first coordinates, which are all zero.
We thus define the \emph{well diagram} as the multiset of points $u_i$
with multiplicities $\mu_i$, denoting it as $\Udgm{\Ugroup}$.
For technical reasons that will become obvious in the next section,
we add infinitely many copies of $0$ to this diagram.
Hence, each point in $\Udgm{\Ugroup}$ is either $0$, a positive real number,
or $\infty$, and the diagram itself is a multiset of points
on the extended line, $\bar{\Rspace} = \Rspace \cupsp \{ \pm \infty \}$.
It has infinitely many points at $0$ and a finite number of non-zero points.

As suggested by the heading of this section, we think of each point
in the diagram as a measure for how resilient a homology class
of $f^{-1} (\Aspace)$ is against perturbations in $\Pert$.
At each well group $\Ugroup_i$, an entire set of homology classes
falls ill, and we call $u_i$ the \emph{robustness} of each
class $\alpha$ in this set, denoting it as $\Rmeasure{\alpha} = u_i$.

\section{Proving Stability}
\label{sec4}

We are interested in relating the distance between mappings
to the distance between their well diagrams, both defined using
a common perturbation space $\Pert$.
After quantifying these distances, we connect parallel well modules
to form new modules, and we finally prove that the well diagram is stable.

\paragraph{Distance between functions.}
Let $f, g: \Xspace \to \Yspace$ be two admissible mappings between
topological spaces, $\Aspace \subseteq \Yspace$ a subspace,
and $\Pert$ a metric space of perturbations that contains both $f$ and $g$.
Using $\Aspace$, we get the two well functions:
$f_{\Aspace},g_{\Aspace} : \Xspace \to \Rspace$.
The distance between them is the supremum difference
between corresponding values:
\begin{eqnarray*}
  \Maxdist{f_{\Aspace}}{g_{\Aspace}}  &=&
     \sup_{x \in \Xspace} \mid f_{\Aspace}(x) - g_{\Aspace}(x) \mid.
\end{eqnarray*}
The distance between the two well functions is related
to the distance, $\Pertdist{f}{g}$, between the two mappings
$f$ and $g$ in $\Pert$.
Specifically, the distance between the well functions cannot
exceed the distance between the mappings.
\begin{result}[Distance Lemma]
  Assuming the above notation, we have
  $\Maxdist{f_{\Aspace}}{g_{\Aspace}} \leq \Pertdist{f}{g}$.	
\end{result}
\proof
 Fix a point $x \in \Xspace$ and let $r = f_{\Aspace}(x)$.
 By construction of $f_{\Aspace}$, there exists an $r$-perturbation
 $h$ of $f$ in $\Pert$ with $x \in h^{-1}(\Aspace)$.
 By the triangle inequality, $\Pertdist{g}{h} \leq \Pertdist{f}{g} + r$. 
 Hence, $g_{\Aspace}(x) \leq r + \Pertdist{f}{g}$.
 Thus, $\mid f_{\Aspace}(x) - g_{\Aspace}(x) \mid \leq \Pertdist{f}{g}$.
\eop

\paragraph{Distance between diagrams.}
Let $\Ggroup (r)$ be the homology group and $\Vgroup (r) \subseteq \Ggroup (r)$
the well group of $g_\Aspace^{-1} [0,r]$.
As for $f$, we insert quotients between contiguous well groups
and connect them with forward and backward maps to form a
well module, denoted as $\Vgroup$.
The corresponding well diagram, $\Udgm{\Vgroup}$,
is again a multiset of points in $\bar{\Rspace}$,
consisting of infinitely many copies of $0$ and
finitely many non-zero points.
Recall that the bottleneck distance between the diagrams
of $f$ and $g$ is the length of the longest edge in the minimizing matching.
Because our diagrams are one-dimensional, the bottleneck
distance is easy to compute.
To describe the algorithm, we order the positive points
in both diagrams, getting
$$
\begin{array}{ccccccccc}
  0 & \leq & u_1 & \leq & u_2 & \leq & \ldots & \leq & u_M ;  \\
  0 & \leq & v_1 & \leq & v_2 & \leq & \ldots & \leq & v_M,
\end{array}
$$
where we add zeros to make sure we have two sequences of the same length.
The \emph{inversion-free matching} pairs $u_i$ with $v_i$ for all $i$.
We prove that this matching gives the bottleneck distance.
\begin{result}[Matching Lemma]
  Assuming the above notation, the bottleneck distance
  between $\Udgm{\Ugroup}$ and $\Udgm{\Vgroup}$ is equal to
  $\max_{1 \leq i \leq M} |u_i - v_i|$.
\end{result}
\proof
 For a given matching, we consider the vector of absolute differences,
 which we sort largest first.
 Comparing two such vectors lexicographically,
 we now prove that the inversion-free matching gives the minimum vector.
 This implies the claimed equality,
 \begin{eqnarray*}
   \Bdist{\Udgm{\Ugroup}}{\Udgm{\Vgroup}}  &=&  \max_{1 \leq i \leq M} |u_i - v_i| ,
 \end{eqnarray*}
 To prove minimality, we consider a matching that has at least one inversion,
 that is, pairs $(u_i, v_t)$ and $(u_j, v_s)$ with $i < j$ and $s < t$.
 If $u_i = u_j$ or $v_s = v_t$ then switching to the pairs
 $(u_i, v_s)$ and $(u_j, v_t)$ preserves the sorted vector of
 absolute differences.
 Otherwise, the new vector is lexicographically smaller than the old vector.
 Indeed, the minimum of the four points is $u_i$ or $v_s$
 and the maximum is $u_j$ or $v_t$.
 If the minimum and the maximum are from opposite diagrams then
 they delimit the largest of the four absolute differences,
 and this largest difference belongs to the old vector.
 Otherwise, both absolute differences shrink when we switch the pairs.
 Repeatedly removing inversions as described eventually leads to the
 inversion-free matching,
 which shows that it minimizes the vector and its largest entry
 is the bottleneck distance.
\eop

\paragraph{Bridges.}
The main tool in the proof of stability is the concept of a short bridge
between parallel filtrations.
The length of these bridges relates to the distance between the functions
defining the filtrations.
Let $\ee = \Pertdist{f}{g}$.
By the Distance Lemma, we have $\Maxdist{f_\Aspace}{g_\Aspace} \leq \ee$,
which implies that the sublevel set of $g_\Aspace$ for radius $r$
is contained in the sublevel set of $f_\Aspace$ for radius $r+\ee$.
Hence, there is a homomorphism $\bridge_r : \Ggroup (r) \to \Fgroup (r+\ee)$,
which we call the \emph{bridge} from $\Ggroup$ to $\Fgroup$ at radius $r$.
We use the bridge to connect the initial segment of $\Ggroup$ to the terminal
segment of $\Fgroup$.
The endpoints of the bridge satisfy the property expressed in the
Shrinking Wellness Lemma.
\begin{result}[Bridge Lemma]
  Let $\bridge_r: \Ggroup (r) \to \Fgroup (r+\ee)$ be the bridge at $r$,
  where $\ee = \Pertdist{f}{g}$ and $f,g \in \Pert$.
  Then $\Ugroup (r+\ee) \subseteq \bridge_r (\Vgroup (r))$.
\end{result}
\proof
 Let $\alpha$ be a homology group in $\Ugroup (r+\ee)$.
 By definition of the well group, $\alpha$ belongs to the image of
 $\Hgroup (h^{-1} (\Aspace))$ in $\Fgroup (r+\ee)$
 for every $(r+\ee)$-perturbation $h$ of $f$ in $\Pert$.
 This includes all $r$-perturbations of $g$.
 It follows that the preimage of $\alpha$ in $\Ggroup (r)$ belongs
 to the well group, that is, $\bridge_r^{-1} (\alpha) \in \Vgroup (r)$.
\eop

Everything we said about bridges is of course symmetric in $\Fgroup$ and $\Ggroup$.
In other words, $f_\Aspace^{-1} [0,r] \subseteq g_\Aspace^{-1} [0,r+\ee]$
and there is a bridge from $\Fgroup (r)$ to $\Ggroup (r+\ee)$
for every $r \geq 0$.

\begin{figure}[hbt]
 \vspace{0.1in}
 \centering
 \resizebox{!}{1.2in}{\input{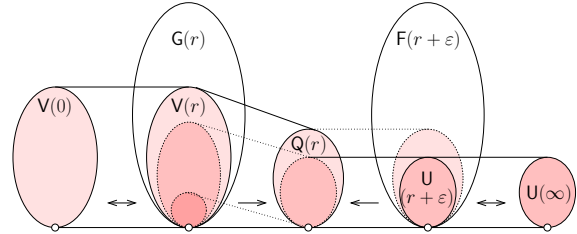}}
  \caption{The zigzag module obtained by connecting an initial segment
   of $\Vgroup$ to a terminal segment of $\Ugroup$.}
 \label{fig:bridge}
\end{figure}

\paragraph{New modules.}
We use the Bridge Lemma to construct new zigzag modules from the well modules of $f$ and $g$.
Specifically, we use $\bridge_r$ to connect the initial segment of $\Vgroup$,
from $\Vgroup (0)$ to $\Vgroup (r)$, to the terminal segment of $\Ugroup$,
from $\Ugroup (r+\ee)$ to $\Ugroup (\infty)$.
To complete the module, we insert
$\Qgroup (r) = \Vgroup (r) / (\Vgroup (r) \capsp \kernel{\bridge_r})$
between $\Vgroup (r)$ and $\Ugroup (r+\ee)$.
The forward map, from $\Vgroup (r)$ to $\Qgroup (r)$, is surjective,
and the backward map, from $\Ugroup (r+\ee)$ to $\Qgroup (r)$ is injective;
see Figure \ref{fig:bridge}.
The new zigzag module is thus of the same type as the well modules
implying it has a left filtration basis that gives rise to a family
of compatible bases for the groups in the left filtration.

\begin{figure}[hbt]
  \vspace{0.1in}
  \centering
  \resizebox{!}{1.3in}{\input{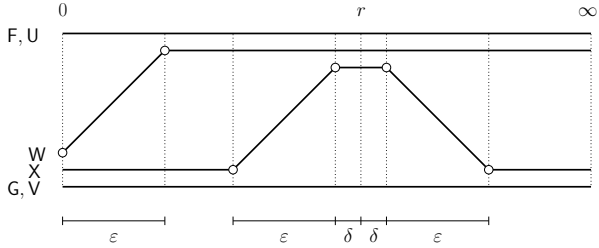}}
   \caption{The four curves represent four filtrations as well as
    four the zigzag modules.
    The middle two are constructed from the outer two by adding bridges
    connecting the dots.}
  \label{fig:newmodules}
 \end{figure}

A particular construction starts with the filtrations
$\Fgroup (0) \to \ldots \to \Fgroup (\infty)$ and
$\Ggroup (0) \to \ldots \to \Ggroup (\infty)$
and adds $\bridge_0: \Ggroup (0) \to \Fgroup (\ee)$.
Following the bridge from $\Ggroup$ to $\Fgroup$ at $0$,
we get a new filtration and a new zigzag module, denoting the latter as $\Wgroup$;
see Figure \ref{fig:newmodules}.
The decomposition of $\Wgroup (0) = \Vgroup (0)$ by the left filtration of $\Wgroup$
is similar to the decomposition of $\Ugroup (0)$ by the left filtration of $\Ugroup$;
see Figure \ref{fig:left-filtration}.
Letting $i$ be the index such that $u_i \leq \ee < u_{i+1}$,
we have $\Ugroup (\ee) = \Ugroup_i$.
The classes in $\Agroup_{i-1}$ and in $\Ugroup_0 / \Bgroup_i$ die before
we reach $\Fgroup (\ee)$.
The remaining classes form
$\Ugroup (\ee) \simeq \Bgroup_i / \Agroup_{i-1}$.
Correspondingly, there are homology classes in $\Wgroup (0)$ that die before
we reach $\Fgroup (\ee)$,
namely the ones in the kernel of the forward map, from $\Wgroup (0)$ to $\Qgroup (0)$,
and in the preimage of the cokernel of the backward map, from $\Ugroup (\ee)$ to $\Qgroup (0)$.
The remaining classes form 
$\Wgroup (\ee) \simeq \bridge_0^{-1} (\Ugroup (\ee))
                    / (\Wgroup (0) \capsp \kernel{\bridge_0})$.
The two quotient groups, $\Ugroup (\ee)$ and $\Wgroup (\ee)$,
are decomposed in parallel so that choosing a basis for $\Ugroup (\ee)$
gives one for $\Wgroup (\ee)$.
This will be useful shortly.
  
\paragraph{Main result.}
We are now ready to state and prove the stability of the well diagram.
\begin{result}[Stability Theorem for Well Diagrams]
  Let $f$ and $g$ be admissible mappings between topological spaces $\Xspace$ and $\Yspace$,
  let $\Aspace$ be a subspace of $\Yspace$,
  and let $\Pert$ a metric space of perturbations containing both $f$ and $g$.
  Then
  $\Bdist{\Udgm{\Ugroup}}{\Udgm{\Vgroup}} \leq \Pertdist{f}{g}$
  for the well modules $\Ugroup$ and $\Vgroup$ of $f$ and $g$.
\end{result}
\proof
 We construct a bijection from $\Udgm{\Ugroup}$ to $\Udgm{\Vgroup}$
 such that the difference between matched points is at most
 $\ee = \Pertdist{f}{g}$.
 Specifically, we match each point $u \leq \ee$ in $\Udgm{\Ugroup}$ with a
 copy of $0$ in $\Udgm{\Vgroup}$,
 and we use the parallel bases of $\Ugroup (\ee)$ and $\Wgroup (\ee)$ for the rest,
 where $\Wgroup$ is the zigzag module obtained by adding the bridge from
 $\Ggroup$ to $\Fgroup$ at radius $0$, as described above.

 Let $\alpha$ belong to the left filtration basis of $\Ugroup (0)$
 such that its image belongs to the basis of $\Ugroup (\ee)$.
 Let $r$ be the value at which $\alpha$ falls ill and note that $r > \ee$.
 Let $\beta$ belong to the left filtration basis of $\Vgroup (0) = \Wgroup (0)$
 such that the images of $\alpha$ and $\beta$ in $\Wgroup (\ee) = \Ugroup (\ee)$ coincide.
 We now construct yet another zigzag module,
 by adding a first bridge from $\Ggroup (r-\ee-\dd)$ to $\Fgroup (r-\dd)$
 and a second bridge from $\Fgroup (r+\dd)$ back to $\Ggroup (r+\ee+\dd)$,
 where $\dd > 0$ is sufficiently small such that there is no death
 in the interval $[r-\dd, r+\dd]$, except possibly at $r$.
 We denote the resulting module by $\Xgroup$;
 see Figure \ref{fig:newmodules}.
 We note that all maps between groups are induced by inclusions so that the
 diagram formed by the filtrations and the bridges between them commutes.

 By construction, the image of $\beta$ in $\Fgroup (r-\dd)$ is non-zero
 and belongs to $\Ugroup (r-\dd)$.
 In contrast, the image of $\beta$ in $\Fgroup (r+\dd)$ is either zero
 or lies outside $\Ugroup (r+\dd)$.
 Applying the Bridge Lemma going backward along the first bridge,
 we note that the image of $\beta \in \Wgroup (0) = \Xgroup (0)$
 in $\Ggroup (r-\ee-\dd)$ is non-zero and belongs to $\Vgroup (r-\ee-\dd)$.
 Applying the Bridge Lemma going forward along the second bridge,
 we note that the image of $\beta$
 in $\Ggroup (r+\ee+\dd)$ is either zero or lies outside $\Vgroup (r+\ee+\dd)$.
 Since we can choose $\dd > 0$ as small as we like,
 this implies that $\beta$ falls ill somewhere in the interval
 $[r-\ee, r+\ee]$.
 In the matching, this radius is paired with $r$, the radius at which $\alpha$
 falls ill in $\Ugroup$.
 The absolute difference between the two radii is at most $\ee$,
 as required.
\eop

\section{Applications}
\label{sec5}

In this section, we use the stability of the transversality measure
to derive stability results for fixed points, periodic orbits,
and apparent contours.
All three problems can be recast in terms of intersections between
topological spaces and are therefore amenable to the tools 
developed in this paper.

\paragraph{Fixed points.}
A \emph{fixed point} of a continuous mapping from a topological space
to itself is a point that is its own image.
Assuming this space is the $m$-dimensional Euclidean space
and $b$ is the mapping,
we introduce a mapping $f: \Rspace^m \to \Rspace^m$ defined
by $f(x) = x - b(x)$.
A fixed point of $b$ is a root of $f$, that is, $f(x) = 0$.
Writing $\Xspace = \Yspace = \Rspace^m$ and $\Aspace = \{0\}$,
the origin of $\Rspace^m$,
we get the setting studied in this paper.
Each fixed point $x$ of $b$ corresponds to a class in the
$0$-dimensional homology group of $f^{-1} (0)$.
Using the methods of this paper, we assign a non-negative
robustness measure, $\Rmeasure{x}$, to $x$.
It gives the magnitude of the perturbation necessary to remove this
fixed point.
This does not mean that a perturbation of smaller magnitude
has a fixed point at precisely the same location
but rather that it has one or more fixed points in lieu of $x$.
Letting $\Rmeasure{x}$ be the maximum robustness
of all the fixed points of $f$, 
every perturbation of magnitude less than $\Rmeasure{x}$
has at least one fixed point.
This implication suffices to give a new proof of
a classic topological result on fixed points;
see \cite{Hat02}.
Let $\Bspace^m$ be the closed unit ball in $\Rspace^m$.
\begin{result}[Brouwer's Fixed Point Theorem]
  Every continuous mapping $b : \Bspace^m \to \Bspace^m$
  has a fixed point.
\end{result}
\proof
 Extend $b$ to a mapping from $\Rspace^m$ to $\Rspace^m$ by
 defining $b(x)$ equal to its value at $x / \|x\|_2$
 whenever $x \not\in \Bspace^m$.
 Let $f: \Rspace^m \to \Rspace^m$ be defined by $f(x) = x - b(x)$
 and let $g: \Rspace^m \to \Rspace^m$ be the identity,
 defined by $g(x) = x$.
 We may assume that $f$ is admissible, else the homology group
 of $f^{-1} (0)$ has infinite rank and $f$ has infinitely many roots.
 The other mapping, $g$, is clearly admissible, with a single root
 at $x = 0$.
 Letting $\Pert$ be the space of all continuous mappings
 from $\Rspace^m$ to $\Rspace^m$,
 and measuring the distance between $f$ and $g$ by taking the
 supremum of the Euclidean distance between corresponding image points,
 we get
  \begin{eqnarray*}
   \Pertdist{f}{g}  &=&  \sup_{x \in \Rspace^m} \Edist{f(x)}{g(x)} \\
                   &=&  \sup_{x \in \Rspace^m} \|b(x)\|_2 ,
 \end{eqnarray*}
 which is at most $1$.
 The well diagram of the identity consists of a single, non-zero
 point at plus infinity.
 The Stability Theorem for Well Diagrams implies that
 the well diagram of $f$ also has a point at plus infinity.
 But this implies that $f$ has a root and, equivalently,
 that $b$ has a fixed point.
\eop

The above reduction of fixed points to an intersection setting uses the
difference between two points,
an operation not available if the mapping $b: \Mspace \to \Mspace$
is defined on a general metric space.
In this case, we can use the correspondence between the fixed points of $b$
and the intersection points between the graph of $b$ and the diagonal
in $\Mspace \times \Mspace$.
To apply the results of this paper, we set
$\Xspace = \Mspace$, $\Yspace = \Mspace \times \Mspace$,
and $\Aspace = \{ (x',x') \mid x' \in \Mspace \}$.
Furthermore, we define the distance between two points
$x = (x', x'')$ and $y = (y', y'')$ in $\Mspace \times \Mspace$ equal to
\begin{eqnarray*}
  \Ydist{x}{y}  &=&  \left\{\begin{array}{cl}
                       \infty            &  \mbox{\rm if~} x' \neq y' ; \\
                       \Mdist{x''}{y''}  &  \mbox{\rm if~} x' = y' .
                     \end{array}\right.
\end{eqnarray*}
We restrict $\Pert$ to those mappings 
$h: \Mspace \to \Mspace \times \Mspace$ that arise as graphs of a continuous
mapping from $\Mspace$ to itself.
It is not difficult to see that this setting gives the same robustness
values for the case $\Mspace = \Rspace^m$ discussed above.

\paragraph{Periodic orbits.}
We generalize the above setting by allowing for fixed points of iterations
of the mapping.
Letting $\Mspace$ be a metric space and $f: \Mspace \to \Mspace$
a mapping, we write $f^j: \Mspace \to \Mspace$
for the $j$-fold composition of $f$ with itself.
A sequence
\begin{eqnarray*}
  \seq{j}{x}  &=&  (x, f(x), f^2(x), \ldots, f^{j-1}(x))
\end{eqnarray*}
is an \emph{order-$j$ periodic orbit} of $f$ if $f^j (x) = f \circ f^{j-1} (x) = x$.
It is straightforward to see the following relationship between
$f$ and its $j$-fold composite.
\begin{result}[Orbit Lemma]
  A point $x \in \Mspace$ is a fixed point of $f^j$ iff
  $\seq{j}{x}$ is an order-$j$ periodic orbit of $f$.
\end{result}
We can therefore use the methods of this paper to measure the robustness
of $x$, that is, to determine how much $f^j$ needs to be perturbed
to remove the fixed point.
However, it is more interesting to measure how much $f$ needs to be perturbed
to remove the periodic orbit.
This is different because not every mapping can be written as the $j$-fold
composite of another mapping.
This motivates us to introduce $\Pert$ as the space of perturbations
of $f^j$ that are $j$-fold composites of perturbations $h$ of $f$.
Using this space $\Pert$,
we intersect the images of the homomorphisms induced by the $h^j$.
With this setup, we construct the well diagram of $f^j$
and interpret the resulting values as the robustness of the
order-$j$ periodic orbits of $f$.

\paragraph{Apparent contours.}
As mentioned in the introduction, \cite{EMP09} reduces the stability of the
contour of a mapping to the stability of well diagrams, the main result of this paper.
We briefly review the reduction.
Let $\Mspace$ be a compact, orientable $2$-manifold and
$f: \Mspace \to \Rspace^2$ a smooth mapping.
The derivative of $f$ at a point $x$ is a linear map from the tangent
space to $\Rspace^2$.
The point $x$ is \emph{critical} if the derivative at $x$ is not surjective,
and the \emph{apparent contour} of $f$ is the set of images of critical points.
Beyond smoothness of $f$, we assume that the well functions 
it defines are admissible.
Specifically, for each $a \in \Rspace^2$,
the function $f_a: \Mspace \to \Rspace$ is defined by mapping every point $x$
to $f_a (x) = \Edist{f(x)}{a}$ and we assume that $f_a^{-1} (0)$
consists of a finite number of points.

To study the apparent contour, we consider the entire $2$-parameter
family of well functions.
Fixing a value $a \in \Rspace^2$,
the sublevel sets of $f_a$ form a filtration of homology groups
and a zigzag module of well groups.
Each point in the preimage of $a$ falls ill at a particular radius
interpreted as the robustness of that point.
The main result of this paper implies that this measure is stable,
that is,
$\Bdist{\Udgm{\Ugroup}}{\Udgm{\Vgroup}}  \leq  \Maxdist{f_a}{g_a}$,
where $\Ugroup$ and $\Vgroup$ are the well modules defined by the
mappings $f, g: \Mspace \to \Rspace$ and the value $a \in \Rspace^2$.
As shown in \cite{EMP09}, this implies that the apparent contours
of $f$ and of $g$ are close.
The sense in which they are close is interesting in its own right
and we refer to that paper for details.

\section{Discussion}
\label{sec6}

The main contribution of this paper is the definition of a robustness
measure for the homology of the intersection between topological spaces,
and a proof that this measure is stable.
While robustness and persistence are related, there are also
differences between these notions.
For example, robustness adapts to a given metric space of perturbations,
and this extra degree of freedom is sometimes essential,
such as for a meaningful analysis of periodic orbits.
The results in this paper
raise a number of questions and invite extensions in several directions.
\begin{itemize}\denselist
  \item Fixed points of mappings play an important role in game theory
    \cite{vSvS08}.
    Can the results of this paper be used to gain insights into
    the nature of fixed points as they arise in different games?
    What are contexts in which the robustness of a fixed
    point is relevant to the understanding of the dynamics of a game?
  \item The three applications sketched in Section \ref{sec5} barely
    scratch the surface of the possible.
    An interesting direction for further research are mappings
    from lower to higher dimensions.
    For example, the boundary of a computer-aided design model is the image
    of a mapping from a $2$-manifold to $\Rspace^3$.
    Can our results be used to detect and remove accidental self-intersections,
    a problem of significant economic importance \cite{GCCBJ01}.
  \item Except for a few special settings, we have no
    algorithms for computing well diagrams.
    The main obstacle is the possibly infinite set of perturbations 
    that appears in the definition of well groups.
    However, since the groups that arise for admissible mappings
    are finite, only a finite number of perturbations are relevant.
    Can we approach the algorithmic question from this direction?
\end{itemize}

\newpage

\end{document}